\documentclass[useAMS,usenatbib,caption]{mnras} %
\usepackage{epsfig}
\usepackage[update,prepend]{epstopdf}
\usepackage{natbib}
\usepackage{textcase}
\usepackage{amsmath}	
\usepackage{amssymb}
\usepackage{cuted}
\usepackage{flushend}

\usepackage[T1]{fontenc}

\usepackage{hyperref}
\hypersetup{colorlinks=true,
	linkcolor=blue,
	filecolor=magenta,      
	urlcolor=cyan,}

\usepackage{xcolor}

\usepackage{ulem} 

\newcommand\gaia{{\it Gaia}}
\newcommand\hipparcos{{\it Hipparcos}}

\def \araa            {{\it Annu. Rev. Astron. Astrophys.}}
\def \aj              {{\it Astron. J.}}

\def \aap             {{\it Astron. Astrophys.}}

\def \apj             {{\it Ap. J.}}
\def \apjl            {{\it Ap. J. Lett.}}

\def \apjs            {{\it Ap. J., Supp. Ser.}}

\def \mnras           {{\it MNRAS}}

\title[Triage of Astrometric Binaries]
{Triage of Astrometric Binaries---how to find triple systems and dormant black-hole 
secondaries in the \gaia\ orbits
}
\author[Shahaf et al.]
{S.\ Shahaf$^1$,  T.\ Mazeh$^1$, S. \ Faigler$^1$, B.\ Holl$^2$
\\
$^1$School of Physics and Astronomy, 
Raymond and Beverly Sackler Faculty of Exact Sciences,\\ 
Tel Aviv University, Tel Aviv  69978, Israel\\
$^2$Astronomical Observatory, University of Geneve, 51 ch. des Maillettes, 1290 Versoix, Switzerland
}

\pagerange{\pageref{firstpage}--\pageref{lastpage}} \pubyear{2007}

\def\LaTeX{L\kern-.36em\raise.3ex\hbox{a}\kern-.15em
    T\kern-.1667em\lower.7ex\hbox{E}\kern-.125emX}

\begin{document}
\label{firstpage}
\maketitle

\date{Received / Accepted }
\pagerange{\pageref{firstpage}--\pageref{lastpage}} \pubyear{2007}

%
%

\begin{abstract} 
Preparing for the expected wealth of  \gaia\ detections, 
we consider here a simple algorithm for classifying unresolved astrometric binaries with main-sequence (MS) primary
into three classes: 
binaries with a probable MS secondary, with two possible values for the mass ratio;
probable hierarchical triple MS systems with an astrometric secondary as a close binary, with a limited range of mass-ratio values; and binaries with a compact-object secondary, with a minimal value of the mass ratio.
This is done by 
defining  a unit-less observational parameter 'Astrometric
Mass-Ratio Function' (AMRF),  $\mathcal{A}$, of a binary, based on primary-mass estimation, in addition to the astrometric parameters---the angular semi-major axis, the period and the parallax. 
We derive the $\mathcal{A}$ value that differentiates the three classes by forward modeling representative binaries of each class, 
assuming some mass-luminosity relation.
To demonstrate the potential of the algorithm, we consider the orbits of 98 \hipparcos\ astrometric binaries with main-sequence primaries, 
using the \hipparcos\ parallaxes and the primary-mass estimates. 
For systems with known spectroscopic orbital solution, 
our results are consistent with the spectroscopic elements, validating the suggested approach.
The algorithm will be able to identify hierarchical triple systems and {\it dormant} neutron-star and black-hole companions in the \gaia\ astrometric binaries.
\end{abstract}


\section{Introduction}        
\label{introduction}                            

The \gaia\ mission 
\citep{gaia16,gaia18}
carries the potential of discovering a large number of wide binaries through the detection of their astrometric motion \citep{lindegren16,lindegren18}. 
The expected large sample of astrometric binaries will allow studying  the statistical characteristics of the wide-binary population in detail, and compare them to the short-period binaries, which are mostly studied by observing spectroscopic and eclipsing systems. 

One aspect of the stellar wide binary population is the frequency of hierarchical triple systems, for which the secondary (or the primary) of the astrometric binary is by itself a short-period binary
\citep[e.g.,][]{tokovinin06,tokovinin18}.
%
Another interest in the upcoming large sample of \gaia\ astrometric binaries stems from the capability of the mission to detect binaries with {\it dormant} black hole (BH) companions. In these systems the BH is not accreting material, because of the relatively large separation between the two components, and therefore does not emit any X-rays. Presumably, most of the BH binaries are in their dormant state, waiting to be discovered by \gaia\ \citep{breivik17, mashian17,piran17,yalinewich18,kinu18}. \gaia\ will detect the large  astrometric motion of the optical companions of these binaries. Obviously, their mass ratios, from which one can derive the BH masses, provided the primary masses can be estimated, is of crucial importance \citep[e.g.,][]{farr11,blackcat16}.

Preparing for the wealth of wide binaries expected from \gaia\ DR3 and 
DR4,\footnote{https://www.cosmos.esa.int/web/gaia/release}
 we consider here a simplistic first-order approach to identify hierarchical triple systems and binaries with compact-object companions. 

\gaia\ will not be able to resolve most of its astrometric binaries.
 As known since the very early days of astrometry \citep[see, for example, an historical review by][]{vandekamp75}, 
the key issue for the  analysis of such systems is the relation between the observed semi-major axis of the photo-center and that of the primary. 

It is also known that if the primary and secondary are both on the main sequence (MS), one can bypass the problem by assuming some mass-luminosity law that applies to the primary and secondary alike
\citep[e.g.,][]{ren13}. 
Building on this idea, we present here a first-order approach to constrain the nature of the secondary, assuming the mass of its primary component can be estimated. Although the discussion is extremely simplified, it lays out a useful path for the analysis of astrometric binaries, and specifically for the detection of compact-object massive secondaries.

We reformulate the analysis by defining a unit-less observational parameter 'Astrometric Mass-Ratio Function' (AMRF), $\mathcal{A}$, of a binary, based on primary-mass estimation, in addition to the astrometric parameters. This is 
similar to the spectroscopic {\it reduced} mass function of \citet{shahaf17}. We show how one can use this parameter for studying the mass ratio of a binary, provided the primary is an MS star. Additionally, we show how one can use the value of
$\mathcal{A}$
to divide  the population of unresolved astrometric binaries into three observational classes---binaries with a probable MS secondary,
binaries with a secondary probably composed of a close pair of MS stars
and 
binaries that have compact-object secondary.

We derive the $\mathcal{A}$ values that differentiate between the three classes by forward modeling representative binaries of each class, assuming a power-law  mass-luminosity relation.
Using our first-order simplistic approach, one can obtain two possible mass-ratio values for binaries of the first class,  obtain a permitted range of mass-ratio values for binaries of the second class, and determine a minimal mass ratio for binaries of the third class. 

To demonstrate the effectiveness of the proposed approach we apply our technique to a small sample of \hipparcos\ astrometric binaries \citep{lindegren97}, using this time a realistic mass-luminosity relation. Validating the suggested approach, we show that the results of our analysis are consistent with the known spectroscopic orbits of the SB9 catalogue\citep{pourbaix04}.

Sections 2 and 3 present our approach, Section 4 displays our analysis of the \hipparcos\ sample, and Section 5 discusses in short our findings.

\section{The astrometric mass-ratio function of a binary}  
\label{sec:astrometry}      

Consider an astrometric binary, with an orbital separation $a$, orbital period $P$, 
primary and secondary masses $M_1$ and $M_2$, respectively, and mass-ratio, defined as $q=M_2/M_1$. Provided the binary is resolved by the astrometric detector, 
the observed angular semi-major axis of the primary is
\begin{equation}
\alpha_1=
\varpi
\left(\frac{M_1}{M_{\odot}}\right)^{1/3}
\left(\frac{P}{\rm yr}\right)^{2/3}
\frac{q}{(1+q)^{2/3}}
\ ,
\label{eq:semi-alpha}
\end{equation}
where $\varpi$ is the system's parallax.
In case we can estimate the primary mass of the resolved astrometric binary, based on its color and brightness, for example, this equation can be solved for the mass ratio. 

Separating the observationally derived  parameters of the binary, 
we define the Astrometric Mass-Ratio Function, AMRF, similar to the {\it reduced} mass function of spectroscopic binaries, to be
\begin{equation}
\mathcal A_1
=\frac{{\alpha_1}}{\varpi}
\left(\frac{M_1}{M_{\odot}}\right)^{-1/3}
\left(\frac{P}{\rm yr}\right)^{-2/3}\ .
\end{equation}
We then obtain a simple equation
\begin{equation}
\mathcal A_1
=\frac{q}{(1+q)^{2/3}} \ , 
\end{equation}
which can be written as a a third-order polynomial equation, 
\begin{equation}
\mathcal A_1^{-3} q^3-q^2-2q-1=0  \ ,
\label{eq:A1polynomial}
\end{equation}
yielding a unique value of the mass ratio $q$, for all positive finite values of $\mathcal{A}_1$.

In most cases, however, the binary is not resolved by the detector, and the astrometric observations yield only the angular  semi-major axis of the {\it photo-center}. 
In this case we define the photo-centric AMRF, $\mathcal{A}$, as 
%
\begin{equation}
\mathcal{A}
=\frac{{\alpha}}{\varpi}
\left(\frac{M_1}{M_{\odot}}\right)^{-1/3}
\left(\frac{P}{\rm yr}\right)^{-2/3}\ ,
\label{EQ:AMRF}
\end{equation}
%
where $\alpha$ is the observed angular semi-major axis of the 
{\it photo-centric} 
orbit.  

The key issue is the relation between $\alpha$ and 
$\alpha_1$.
 The well-known ratio between the two 
depends on the mass-ratio and the light-ratio of the two components,

\begin{equation}
\alpha=
\alpha_1
\left(1-\frac{\mathcal{S} (1+q)}{q(1+\mathcal{S} )} \right) \ ,
\end{equation}
where $\mathcal S={I_2}/{I_1}$, and $I_1$ and $I_2$ are the intensities of the primary and secondary, integrated over the bandpass of the detector.

We finally get to the Astrometry equation,
%
\begin{equation}
\mathcal A=
\frac{q}{(1+q)^{2/3}}  
\left(1-\frac{\mathcal S (1+q)}{q(1+\mathcal S)} \right) \ ,
\label{EQ:A_vs_q}
\end{equation}
where the second factor reflects the unknown reduction of the {\it observed} astrometric motion around the binary center-of-mass, relative to the motion of the primary.

As one can derive the \gaia\ (or \hipparcos\ for that matter) mass-luminosity relation on the MS, we can obtain the luminosity ratio between any two MS stars. Thus, for a binary with both components on the MS, $\mathcal{S}(q)$ is known for any primary mass. 
We can therefore numerically solve equation~(\ref{EQ:A_vs_q}) for  $q$, based on the observationally derived value of $\mathcal A$. As is well known, in most cases this equation has two roots, as is demonstrated below. 

To discriminate between the different systems
we consider here three representative binaries---two extreme 
and one in-between cases.
One case is a binary with a non-luminous secondary, $\mathcal S=0$, and another is a binary whose primary and secondary are both MS stars. The third, in between case, is an hierarchical MS triple system, where  the astrometric secondary is a close binary by itself.
For such a system we assume that the mass ratio of the close binary is known.
We forward model the three cases, 
{\it assuming}  that the intensity ratio of MS stars, as seen by the astrometric detector, can be approximated as a power law of the form $\mathcal S=q^{\beta}$
(but see below a more realistic approach for the \hipparcos\ binaries). 
We show that these three representative  binaries can define three different observational classes, differentiated by the range of their of possible $\mathcal{A}$ values. Assigning a binary to one of the three classes can then be used to constrain the nature of its  secondary. 

\subsection{A non-luminous secondary}

As detailed above, the case of a non-luminous secondary is quite simple, as
$\mathcal S=0$. We therefore get
\begin{equation}
\mathcal A=
\mathcal A_1
=\frac{q}{(1+q)^{2/3}} \ . 
\label{EQ: A NLC}
\end{equation}
The solution of this equation is the root of the third-order polynomial given in equation~(\ref{eq:A1polynomial}).

\subsection{A MS secondary}

Assuming the two components are on the MS, with simplified mass-luminosity relation 
\begin{equation}
\mathcal{S_{\rm MS}}(q)=
\frac{I_2}{I_1}=
\left(\frac{M_2}{M_1}\right)^{\beta}=
q^{\beta} \ ,
\end{equation}

we get 
%
\begin{equation}
\mathcal A=
\mathcal A_{_{\rm MS}}
=\frac{q}{(1+q)^{2/3}}  
\left(1-\frac{q^{\beta} (1+q)}{q(1+q^{\beta})} \right) \ .
\label{EQ: A2ms}
\end{equation}

\subsection{A secondary composed of two MS stars}
%
If the astrometric secondary is by itself a binary with two  MS stars with mass-ratio $q_2$,
the intensity ratio becomes
%
\begin{equation}
\mathcal{S}_{\rm triple}(q,q_2)=
 \frac{I_{2,A} + I_{2,B}}{I_1}=
q^\beta \frac{1+q_2^\beta}{(1+q_2)^\beta} \ ,
\end{equation}
%
where $I_{2,A}$ and $I_{2,B}$ are the intensities of the close-binary primary and secondary components, respectively. 
The expression for $\mathcal{S}_{\rm triple}(q,q_2)$ has to be inserted into equation~(\ref{EQ:A_vs_q}) for  $q$ and $q_2$. 

The second factor in the expression of $\mathcal{S}_{\rm triple}(q,q_2)$, which only depends on $q_2$, gets its minimum, for  $\beta>1$, at $q_2=1$, with a value of $2^{1-\beta}$. The maximum is attained when $q_2=0$, with a value of unity. This reflects the fact that for a
 given mass of the astrometric secondary, its minimum brightness is reached when the mass of the secondary is divided into two equal halves, whereas the maximum brightness is obtained when all its mass is contained in one star, namely, we are back in the previous case. 
We then get 
%
\begin{equation}
q^\beta \, 2^{1-\beta}
\leq
\mathcal{S}_{\rm triple}(q)
\leq
 q^\beta \ .
\end{equation}

Obviously, the astrometric secondary must be fainter than the primary. In the previous case this is kept by having $q\leq 1$. When the secondary is by itself a binary, this limit depends on $q_2$. The limit is reached when 
\begin{equation}
q\left(\mathcal S_{\rm triple}=1\right) = \big({1+q_2}\big)\cdot{\big(1+q_2^\beta\big)^{-\beta^{-1}}} \ .
\end{equation}
This maximum  of $q$
ranges from $1$, for $q_2=0$ (normal MS case) 
to $2^{(1-\beta^{-1})}$, at $q_2=1$.

\section{The nature of the secondary component}

\label{Sec:NatureOfSec}

\begin{figure*} 
\centering
{\includegraphics[width=.8\linewidth,]{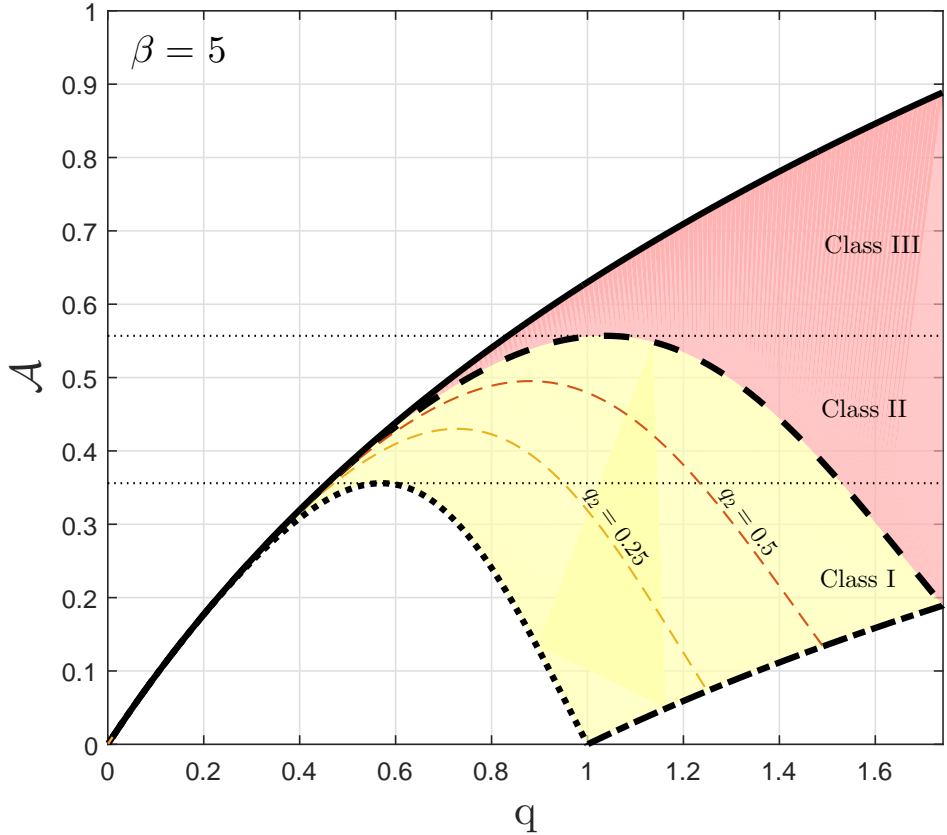}}
\caption{
Astrometric Mass-Ratio Function (AMRF), $\mathcal A$, as a function of $q$ for three cases (see text)---$\mathcal A_1$ (a thick continuous  line for non-luminous secondary), $\mathcal A_{_{\rm MS}}$ (thick dotted line for an MS secondary) 
and  three $\mathcal A_{\rm triple}$ functions.
A thick dashed line for the $\mathcal A_{\rm triple}(q_2=1)$ case, presenting an hierarchical triple MS system, where the astrometric secondary is an equal-mass close binary.
The two thin dashed lines represent hierarchical systems with $q_2=0.25$ (orange) and $q_2=0.5$ (brown).
The thick dashed-dotted line marks the $\mathcal{S}_{\rm triple}=1$ limiting case, where the close-binary astrometric secondary is as bright as the primary (see text).
The functions were derived by assuming a mass-luminosity power law of $L\propto M^{\beta}$ with  $\beta=5$. 
The white area is excluded for any binary or triple system, whereas 
the yellow area can be populated by hierarchical triple MS systems. 
In the light-red area one can find only systems with compact-object secondaries, either as single or as members of a close binary system.	
The thin dotted  horizontal lines separate the diagram according to the different classes  (see text).
}
\label{fig:AMRF}
\end{figure*}

The $\mathcal A$ functions of the three binaries discussed above are plotted in Figure~\ref{fig:AMRF} for $\beta=5$.  
The figure shows that for a non-luminous secondary $\mathcal A_1$ is a monotonically increasing function of the mass ratio $q$. Specifically, for $q\leq1$ we get $\mathcal A_1\lesssim0.6$. 
On the other hand, $\mathcal A_{_{\rm MS}}$ reaches a maximum value of $\sim0.36$ at $q\sim 0.6$, 
then monotonically decreases to a value of zero at $q=1$. 

The  hierarchical triple systems, where the  astrometric secondary which is a binary by itself,
are presented in the figure by three functions, $\mathcal A_{_{\rm triple}}$, with $q_2$ values of 0.25, 0.5 and 1. In fact, the $\mathcal A_{_{\rm MS}}$ function can be viewed as an $\mathcal A_{_{\rm triple}}$ case with $q_2=0$.
 
The  $\mathcal A_{\rm triple}$ functions show similar behavior, starting at zero for $q=0$, attaining a maximum at some value of $q$, depending on $q_2$, and then go down to a value which again depends on $q_2$, as discussed above.

Not surprisingly, for small $q$ all curves coincide, while their largest differences are seen when $q\geq 1$.

%
%
%

When we consider Figure~\ref{fig:AMRF}, we better keep in mind that for any given astrometric binary we observationally obtain only its ordinate value, 
$\mathcal A$. A point in the figure that presents a system  can be  at any  location between the different curves, depending on the light contribution of the secondary, which is unknown. Only the white areas of the figure are excluded. 

The curves of Figure~\ref{fig:AMRF} 
suggest  three classes of astrometric binaries:

\begin{itemize}
\item
Class I---binaries with $\mathcal{A}$ below the maximal value of $\mathcal{A}_{_{\rm MS}}$,  $\max\{\mathcal{A}_{_{\rm MS}}\}$ (0.36 in the figure); most probably binaries with a MS secondary. 
\item
Class II---binaries with $\mathcal{A}$ above  $\max\{\mathcal{A}_{_{\rm MS}}\}$,
 but below $\max\{\mathcal{A}_{_{\rm triple}}\}$ with $q_2=1$ (0.56 in the figure);  probable hierarchical triples, with the astrometric secondary as a close binary composed of two MS stars. 
\item
Class III---binaries with $\mathcal{A}$ above 
$\max\{\mathcal{A}_{_{\rm triple}}\}$;
systems that probably contain a compact object.  
\end{itemize}

The division between the classes is clearly depicted  in Figure~\ref{fig:AMRF}. 

Note that our astrophysical division between the three classes is of a probabilistic nature only. For example, class-I systems can still be hierarchical triple systems (see below), although this is a much less probable scenario. On the other hand, class-II systems cannot be MS binaries. 
Similarly, class-I and -II systems may still have a compact-object secondary, although this is much less probable, whereas class-III binaries must have a massive compact companion. 
Nevertheless, the $\mathcal A$ value of a binary, when compared with the figure, can lead to a first guess of its nature and can help forming a first-order triage of a sample of binaries.

In the next section we apply the suggested triage approach to a small sample of \hipparcos\ MS binaries in order to show how our triage should work in real samples, and  demonstrate its potential.

\section{Analysis of the \hipparcos\ astrometric binaries 
with MS primaries}
\label{sec:hipparcos}

\begin{figure}
 	\centering
 	{		
 		\includegraphics[width=0.95\linewidth,trim={10 0 10 10 },clip]{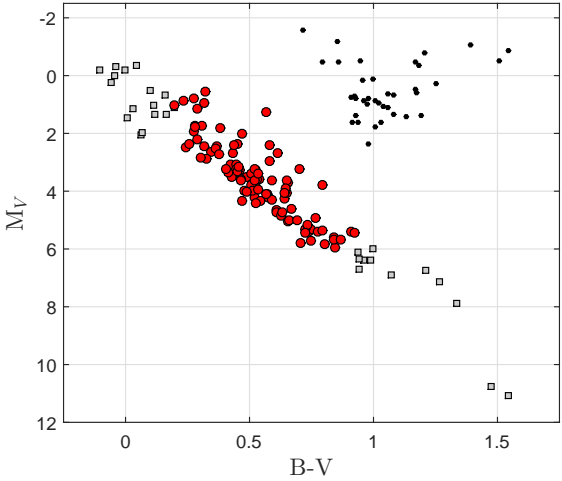}	}
 	\caption{Color-Magnitude Diagram (CMD) of 163 stars with \hipparcos\ astrometric solution within 100 parsecs and $\alpha/\Delta\alpha>3$. Red dots represent the 98 MS stars we considered.
}	
 	\label{fig:CMD}
 \end{figure}

\citet[][]{lindegren97} reported of 235 stars with \hipparcos\ astrometric binary orbits, out of which 163 have 
parallaxes larger than 10 mas and $\alpha_H/\Delta\alpha_H > 3$,
where $\alpha_H$ and $\Delta\alpha_H$ are the \hipparcos\ semi-major orbit axes and their errors.
These systems are plotted on \hipparcos\ Color-Magnitude Diagram (CMD) in Figure~\ref{fig:CMD}, 
using the \hipparcos\ main 
catalogue\footnote{\url{http://cdsarc.u-strasbg.fr/viz-bin/cat/I/239}}
data \citep[][]{ESASP97}, 
which includes the V-band magnitude and the B-V 
 and V-I color indexes.
As we are interested only in MS binaries, we remove from the sample objects that clearly deviate from the MS, based on this CMD, 
coloring in red the remaining 126 objects.

The sample consists mostly of A--K type primaries, at a range of orbital periods that spans between a few weeks and a few decades. The projected semi-major axes
are of 2--200 mas, with the exceptions of Sirius and Procyon that have $\alpha_H$ on the order of $1"$. 

We derived the primary mass, $M_1$, of each of the 126 systems independently from both their B-V and V-I colour indexes.
This was done by interpolating over stellar color and effective temperature values of the table\footnote{Version 2018.05.24, \url{http://www.pas.rochester.edu/~emamajek/EEM_dwarf_UBVIJHK_colors_Teff.txt}} of \citet[][PM13]{pecaut13}.
We used the mean value as our best-estimate mass, and the difference as the uncertainty, with a minimum uncertainty of  $10\%$.

Colour-based estimation of the primary mass of a binary can be affected by the light contribution of the cooler secondary star. 
Simulations we performed for two MS stars using PM13 table showed that the maximum B-V colour shift induced by a secondary is in the range of 0.04--0.06, depending on the primary mass. 
This B-V colour shift induced an underestimation of the primary mass  between  $\sim 0.03 M_{\odot}$ (for $0.8 M_{\odot}$ primary) and  $\sim 0.1 M_{\odot}$ (for $1.8 M_{\odot}$ primary), with very similar values for the V-I colour shift.
We therefore conclude that the secondary light can induces less than $10\%$ error in the primary mass estimate. 

To have an homogeneous sample for our triage we considered only primaries with estimated masses of 0.8--1.8 $\textrm{M}_\odot$, so that the analyzed systems will be within the limits of the  \hipparcos\ mass-luminosity relation we derived below.
We then calculated the  AMRF, $\mathcal{A}$, of each of the remaining 98 systems in the sample according to equation (\ref{EQ:AMRF}) and classified them by our approach.

\subsection{Triage of the \hipparcos\ astrometric binary sample}

 %
 %

The distinction between the three classes of astrometric binaries depends on the maximal values of $\mathcal{A}_{\rm MS}$ and $\mathcal{A}_{\rm triple}$, which depend on the mass-luminosity relation.
We assumed above a simplistic power-law relation, which is far from being realistic. 
In order to apply our algorithm to the \hipparcos\ binaries we fitted the mass-luminosity relation of the \hipparcos\ $H_p$ band by a broken power-law, as described in Appendix~\ref{APPENDIX: Hip M-L relation}. 
The broken power-law induced different shapes to the expected $\mathcal A$ curves, which depend on the primary mass, as demonstrated in Appendix~\ref{APPENDIX: Hip M-L relation}
and figure~\ref{fig:HipparcosAms}. 

\begin{figure}
	\centering
	{		
		\includegraphics[width=1\linewidth,trim={0 0 0 10 },clip]{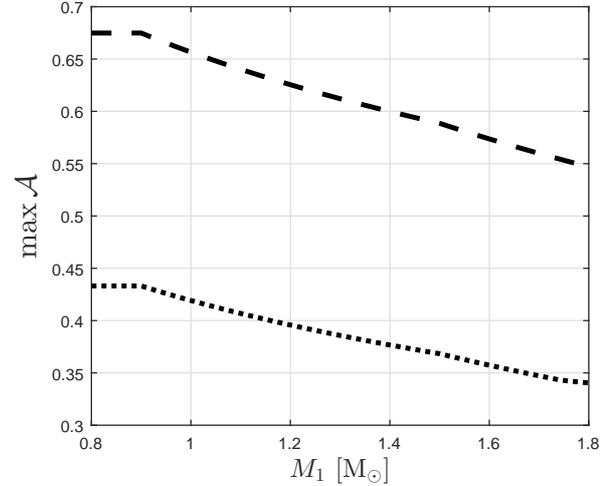}	}
\caption {Maximal value  of $\mathcal{A}_{\rm MS}$ (dotted line) and $\mathcal{A}_{\rm triple}$ (dashed line), as a function of the primary mass, $M_1$, for the \hipparcos\ band. 
}
	\label{fig: Max AMRF}
\end{figure}

We derived $\mathcal{A}$ curves for a range of primary masses, and obtained the  corresponding maximal values of $\mathcal A_{\rm MS}$ and $\mathcal A_{\rm triple}$ as shown in Figure~\ref{fig: Max AMRF}. 
Equipped with these values we then turned to consider the small sample of  \hipparcos\ unresolved astrometric binaries.

%

Our triage process divided the 98 binaries into 3 groups:
\begin{itemize}
\item
68 systems had $\mathcal{A}$ values smaller than the corresponding $\max\{\mathcal{A}_{_{\rm MS}}\}$, and therefore were identified as typical  
class-I binaries. These systems, with probable MS secondaries, had two possible solutions for the mass ratio.

\item
 19 systems had $\mathcal{A}$ values larger, but within $1\sigma$, from their corresponding $\max\{\mathcal{A}_{_{\rm MS}}\}$. These systems were classified as probable class-I binaries, with MS secondaries.  We assigned them a $q$ value that corresponded to their $\max\{\mathcal{A}_{_{\rm MS}}\}$. 

\item
11 systems had $ \mathcal{A}$ values  significantly larger than their corresponding $\max\{\mathcal{A}_{_{\rm MS}}\}$, but not significantly larger than 
$\max\{\mathcal{A}_{_{\rm triple}}\}$, making them typical class-II systems---probably hierarchical triple systems. We assigned them with a range of $q$ values.
\end{itemize}

We have not identified a class-III system which might indicate a compact-object companion.

\subsection{Comparison with the SB9 catalogue} 

In order to validate our approach, we searched the SB9 catalogue\footnote{\url{http://sb9.astro.ulb.ac.be/}} of spectroscopic binaries \citep[][]{pourbaix04} for binaries from   our \hipparcos\ sample. 
We wished to compare the results of the astrometric analysis with the spectroscopic orbital solution.

For single-lined (SB1) systems we derived
 $\textrm{K}_\textrm{ast}$, the expected RV semi-amplitude,  based on the \hipparcos\ orbital elements and our derived $M_1$ and $q$. 
We then compared the expected semi-amplitude with $\textrm{K}_1$, the actual observed semi-amplitude RV reported in SB9. 
For double-lined (SB2) systems, we directly compared the two derived mass ratios. 
The good agreement between our results and the spectroscopic elements showed that even if the secondary substantially contributed to the binary light, our approach yielded a correct mass ratio.

We identified 31 (out of 87) class-I binaries, and 7 (out of 11) suspected hierarchical triples class-II systems in SB9. These 
included 3 class-I and 1 class-II binaries that SB9 identified as systems for which the astrometric {\it primary} is a short-period spectroscopic binary.

\begin{figure*} 
	\centering
	\includegraphics[width=0.85\linewidth,,trim={0 0 0 0 },clip]{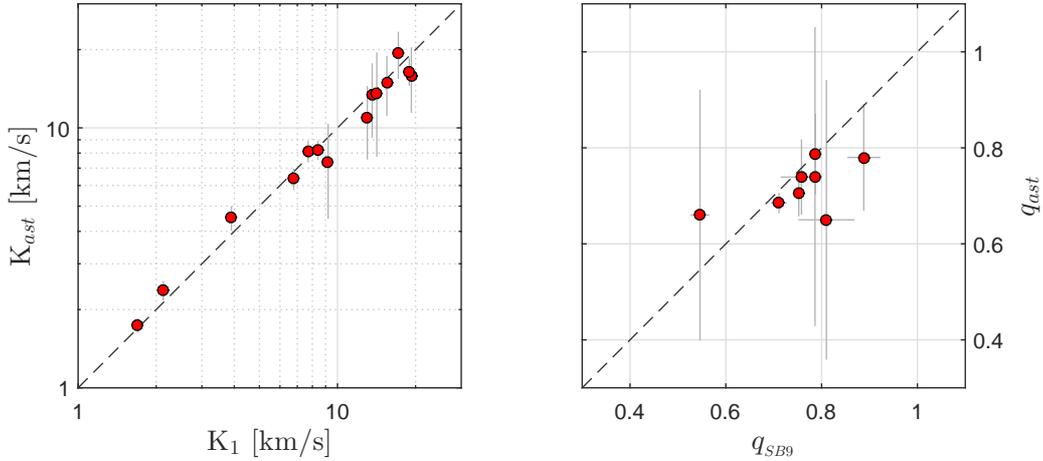}
	\caption{Comparison between 
 the results of the astrometric analysis and the spectroscopic orbital elements of SB9.
{\it Left:} 
Expected RV semi-amplitude, $\textrm{K}_\textrm{ast}$, vs.~the observed RV semi-amplitude, $\textrm{K}_1$,  of 14 class-I systems with SB1 solution (see text).
{\it Right:} 
Mass ratio, $q_\textrm{ast}$, vs.~the observed mass ratio of 8 class-I systems with SB2 solution.
Dashed lines mark the one-to-one relation.}
	\label{fig:KvsK}
\end{figure*}

We first considered the 28 class-I binaries for which the astrometric primary is a single star, 18 of them are SB1s and ten are SB2. However, for four of the SB1s and two of the SB2s we obtained large errors for
$\textrm{K}_\textrm{ast}$  and therefore removed them from our comparison. We were left with only 14 SB1s and eight SB2s for the comparison, the results of which are given in Tables~\ref{table:KvsK} and  \ref{table:qvsq} and 
plotted in Figure~\ref{fig:KvsK}, where 
we chose the $\textrm{K}_\textrm{ast}$ which is closer to the observed  
$\textrm{K}_1$.

%
%
%

As can be seen in the figure, our results are consistent with the spectroscopic elements. Table~\ref{table:qvsq} shows that the average ratio between $q_{\rm ast}$, the mass ratio derived from the astrometric orbit, and $q_{\rm SB9}$ is  $0.96\pm0.12$, consistent with unity. 

Out of the 7 class-II systems published in the SB9 catalogue, 5  were identified or suspected to have a short-period binary secondary \citep[HIP 20935, 61880, 76031, 81023, 93995, see][]{johnson86, latham02,jancart05,griffin13}, confirming our approach. One system is reported to have an equal-mass short-period binary as its primary star \citep[HIP 108478, see][]{bakics10}. 
We have found no reports of statistically significant evidence for higher multiplicity for the remaining system
\citep[HIP 62124, ][]{griffin01}.

\section{Summary and Discussion}        
\label{sec:discussion}                            

Preparing for the expected wealth of  \gaia\ detection, 
we considered here a simple algorithm for classifying unresolved astrometric binaries
into three classes:
\begin{itemize}
\item
Binaries with a probable MS secondary.
\item
Probable hierarchical triple MS systems.
\item
Binaries with a compact-object secondary. 
\end{itemize}

This is done by 
defining  a unit-less observational parameter 'Astrometric
Mass-Ratio Function' (AMRF),  $\mathcal{A}$, of a binary, based on primary-mass estimation, in addition to the astrometric parameters---the angular semi-major axis, the period and the parallax. This is similar in some ways to the spectroscopic modified mass function introduced recently by \citet[][]{shahaf17}. 

The algorithm yields two possible values of the mass ratio
for class-I binaries, if indeed the secondary is a MS star,  a limited range of $q$ values for class-II systems, and a minimal value of $q$ for class-III binaries . 

We note that in actual cases the derivation of the primary mass from the brightness, distance and color of the system should also take into account the contribution of the secondary. This second-order correction can be done iteratively;
in each iteration, the mass ratio found in the previous iteration should be used for a better estimation of the primary mass.
Another possibility that might have an impact on the analysis is that the primary is by itself a close binary, and therefore its mass might not be reflected by the apparent brightness and temperature of the system. We ignored these considerations here, in order to keep the derivation simple. 

In any case, as discussed above, our simulations show that a colour-based mass estimation of the primary is only slightly sensitive to the 
contribution of a MS secondary---the effect is less than $10\%$. This is so because for a MS secondary  to have substantial light contribution, its mass has to be similar to that of the primary, and therefore must have similar colour index. 


To validate our first-order algorithm, we applied it to a small sample of 98 MS \hipparcos\ astrometric binaries, using the \hipparcos\ parallax, absolute magnitude and color. Most of the binaries were found to be class-I systems. We have found 11 class-II systems, 5 of which 
were previously known as such. 
We have not found any class-III system, consistent with the paradigm that neutron-star and BH binaries are quite rare.

The SB9 catalogue includes orbits for 22 systems from our \hipparcos\ sample. We have shown that our analysis is consistent with the orbital elements of SB9 for the SB1 and SB2 binaries alike, provided we choose the pertinent $q$ value. The good agreement with the RV elements demonstrates the potential of our first-order approach.


As emphasized above, the  astrophysical division between the three classes is of one-way non-exclusive probabilistic nature only. Although much less probable, class-I binaries  can still be hierarchical triple systems,  whereas class-II systems cannot have a single MS secondary. 
Similarly, class-I and -II systems may still have a compact-object secondary, although this is much less probable, whereas class-III binaries must have a massive compact companion. Despite its probabilistic nature, the algorithm enables an automated analysis of a sample of astrometric binaries, identifying most of the triple systems and most of the binaries with compact-object secondaries in the sample. 

Given the expected large number of  astrometric binaries to be detected by \gaia, this algorithm can determine the frequency of triple systems \citep[e.g.,][]{mm87,  fuhrmann17, tokovinin06} and the occurrence rate of wide binaries with compact object, BH in particular.

As emphasized recently by \citet{tokovinin18}, his catalog of triple and multiple systems "results from random discoveries and
gives a distorted reflection of the real statistics" of the triple systems. "The volume-limited samples \citep[e.g.,][]{raghavan10} 
are necessarily small and contain only a modest number of
hierarchies, diminishing their statistical value". The inherent limitation of our present knowledge of the multiple-system population is going to be substantially removed by the future \gaia\ releases,
when analyzed by our algorithm. We will be able to find hierarchical triple systems for which the astrometric secondary is a close binary, i.e.,  systems most difficult to identify. The statistical features of this population will shed light on the role of a third stellar distant companion on the formation of close binaries \citep[e.g.,][]{mazeh79, fabrycky07, naoz16}.

Finally, the algorithm can automatically identify \gaia\ binaries with compact-object secondary, BH in particular. 
Enlarging the population of known binaries with dormant BH is crucial for our understanding of how
such systems are formed, including stellar evolution of the BH progenitor and its dependence
on the binarity of the system, the formation of the BH itself with or without a natal kick,
stellar evolution of the companion and its dependence on the proximity of the massive
BH companion and the orbital evolution of the binary, before and after the BH formation
 \citep[e.g.,][]{bel02,repetto17,kochanek18}.
Here again the \gaia\ project will substantially change our comprehension of the formation and evolution of the population of BHs in wide binaries.


\section*{Acknowledgements}

We are indebted to the anonymous referee who helped us focus on the classification aspect of the approach and for very useful and thoughtful suggestions and comments. We deeply thank the \hipparcos\ team for producing such a great sample of astrometric binaries,
Dimitri Pourbaix and Andrei Tokovinin for maintaining the SB9  and the hierarchical system catalogues.
This research was supported by Grant No. 2016069 of the United States-Israel Binational Science Foundation (BSF) and by the Israeli Centers for Research Excellence (I-CORE, grant No. 1829/12).
This work also used the SIMBAD and VIZIER services operated by 
Centre des Donn\'ees Stellaires
(Strasbourg, France) and bibliographic
references from the Astrophysics Data System maintained by
SAO/NASA. 


\bibliographystyle{mnras}
\bibliography{GaiaAstrometry}


%
\begin{appendix}
%

\section{The \hipparcos\ mass-luminosity relation and its resulting AMRF}
\label{APPENDIX: Hip M-L relation}
{
	\begin{figure} 
		\centering
		{\includegraphics[width=0.9\linewidth,trim={0 0 0 0 },clip]{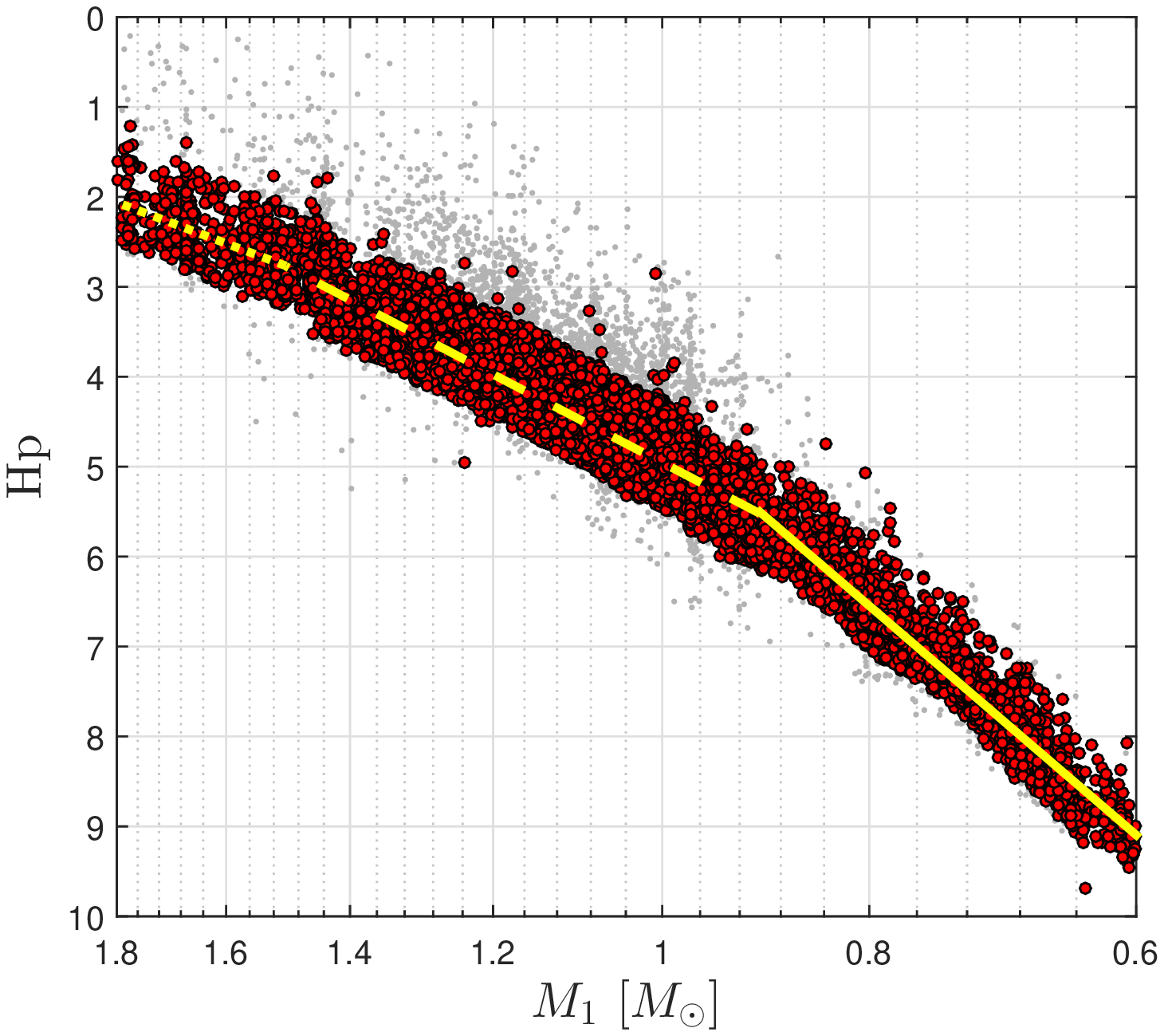}}
		\caption
		{ 
			Mass-Luminosity relation for the sample of \hipparcos\ band. 
			$\textrm{H}_p$ is the absolute magnitude, derived using the parallax estimate by \gaia. The mass, $\textrm{M}_1$, was derived using \citet{pecaut13} MS table and the effective temperature value provided by \gaia.
			Red points are the 10,784 targets that were used for the mass-luminosity fit (yellow lines). Grey points mark the points that were rejected in the last stage due to luminosity excess (see text).
		}
		\label{fig:M-L}
	\end{figure}	

To determine the mass-luminosity relation of the \hipparcos\ band we used the \hipparcos-\gaia\ cross-match catalogue \citep{marrese18} that contained 83,034 entries at the time of our analysis. Out of the entire cross-matched sample, we used $\sim15,000$ \hipparcos\ targets  closer than 100 parsecs, according to their measured \gaia\ parallaxes. 
		
A CMD of the 100 pc \hipparcos-\gaia\ cross-matched sample was constructed using \gaia's parallaxes, G-band magnitudes and colors. Targets that clearly deviate from the MS were manually excluded from the sample, leaving $\sim13,500$ MS candidates. 
 
We restricted the sample to targets that are closer than 100 parsecs and within the 0.6--1.8 $\textrm{M}_\odot$ range. The masses were derived by using the MS tables of \citet{pecaut13} and the effective temperature estimates provided by \gaia\ DR2. These restrictions left a subsample $\sim12,500$ targets.
	
Finally, in order to diminish the impact of binaries on our analysis we compared the bolometric luminosity  as reported in \gaia\ DR2, $ L_{\gaia}$, to the MS luminosity determined by \citet{pecaut13}, $L_{MS}$. Only targets that fulfill
	\begin{equation}
	\frac{\big| L_{\gaia} - L_{MS}\big|} {L_{MS}} < 50\% 
	\end{equation}
	were considered in the analysis. This last step left a total of 10,784 targets in the analyzed sample.
 
	The mass-luminosity relation for the \hipparcos\ 
$\textrm{H}_p$ band was then obtained by fitting a broken power-law to the sample of the form (see Figure~\ref{fig:M-L})
%
	\begin{equation}
	\textrm{H}_p = a\cdot \mathcal{M} + b \ ,
	\label{EQ: MassMag}
	\end{equation}
%
where  $\mathcal{M}=\log {M_1}/{\textsc{M}_\odot}$. The fitted parameters, $a$ and $b$, are given in Table~\ref{table: BrokenPoerLawCoeffs}.  
%
	\begin{table}     
		\centering 
		\begin{tabular}{c c c }
			\hline 
			\hline    
			mass [${\textsc{M}_\odot}$] & $a$ &  $b$    \\
			\hline 
			\hline
			1.5--1.8&$-8.9 \pm 0.6$  & $4.3 \pm 0.1$ \\
			0.9--1.5&$-12.39 \pm 0.08$  & $4.955 \pm 0.006$  \\
			0.6--0.9&$-20.39 \pm 0.09$  & $4.57 \pm 0.01$  \\
			\hline                                        
			\hline                           
\end{tabular}  	
\caption{Fitted parameters for the mass-magnitude relation in equation~(\ref{EQ: MassMag}).  }
		\label{table: BrokenPoerLawCoeffs}  
	\end{table}    
	%

This relation can be expressed in terms of the intensity ratio on the \hipparcos\ band,
	$$\mathcal{S}^{Hp}(M_1,M_2)\equiv 10^{-0.4(\Delta Hp)},$$
where $M_1$, $M_2$ and $\Delta Hp$ are the primary mass, secondary mass, and magnitude difference, respectively. 

We then get for A-type MS stars, of 1.5--1.8 $\textrm{M}_\odot$, a mass-luminosity relation of the form  
	\begin{equation} 
	\mathcal{S}_\textrm{\,A}^{Hp}(q) =\begin{cases} 
	q^{3.6}& 1.5< \frac{q M_1}{\textsc{M}_\odot} < 1.8   ,\\ 
	0.566 \cdot \big(\frac{M_1}{\textsc{M}_\odot}\big)^{1.4} \cdot q^{5}&0.9<  \frac{q M_1}{\textsc{M}_\odot} < 1.5 \ ,\\ 
	0.794 \cdot \big(\frac{M_1}{\textsc{M}_\odot}\big)^{4.6} \cdot  q^{8.2}&  \frac{q M_1}{\textsc{M}_\odot} <0.9 \ ,\\ 
	\end{cases} 
	\label{EQ: S_Hp_A}
	\end{equation}  
for F--G type primaries  of 0.9--1.5 $\textrm{M}_\odot$
	\begin{equation} 
	\mathcal{S}_\textrm{\,FG}^{Hp}(q)=\begin{cases} 
	q^{5}& 0.9< \frac{q M_1}{\textsc{M}_\odot} < 1.5 ,\\ 
	1.40 \cdot  \big(\frac{M_1}{\textsc{M}_\odot}\big)^{3.2}\cdot  q^{8.2}&   \frac{q M_1}{\textsc{M}_\odot} <0.9 \ , \\ 
	\end{cases} 
	\end{equation} 
%
and for K-type primaries of 0.6--0.9 $\textrm{M}_\odot$
	%
	\begin{equation} 
	\mathcal{S}_\textrm{\,K}^{Hp}(q)=
	q^{8.2} \ .
	\end{equation} 

We used these relations to plot AMRF(q) in Figure~\ref{fig:HipparcosAms}  for the three cases---$\mathcal A_1$, $\mathcal A_{_{\rm MS}}$ and $\mathcal A_{\rm triple}$ for $0.8 M_{\odot}$, $1.3 M_{\odot}$ and $1.8 M_{\odot}$ primaries.

%
\begin{figure}
	\centering
	{		
		\includegraphics[width=0.95\linewidth,clip]{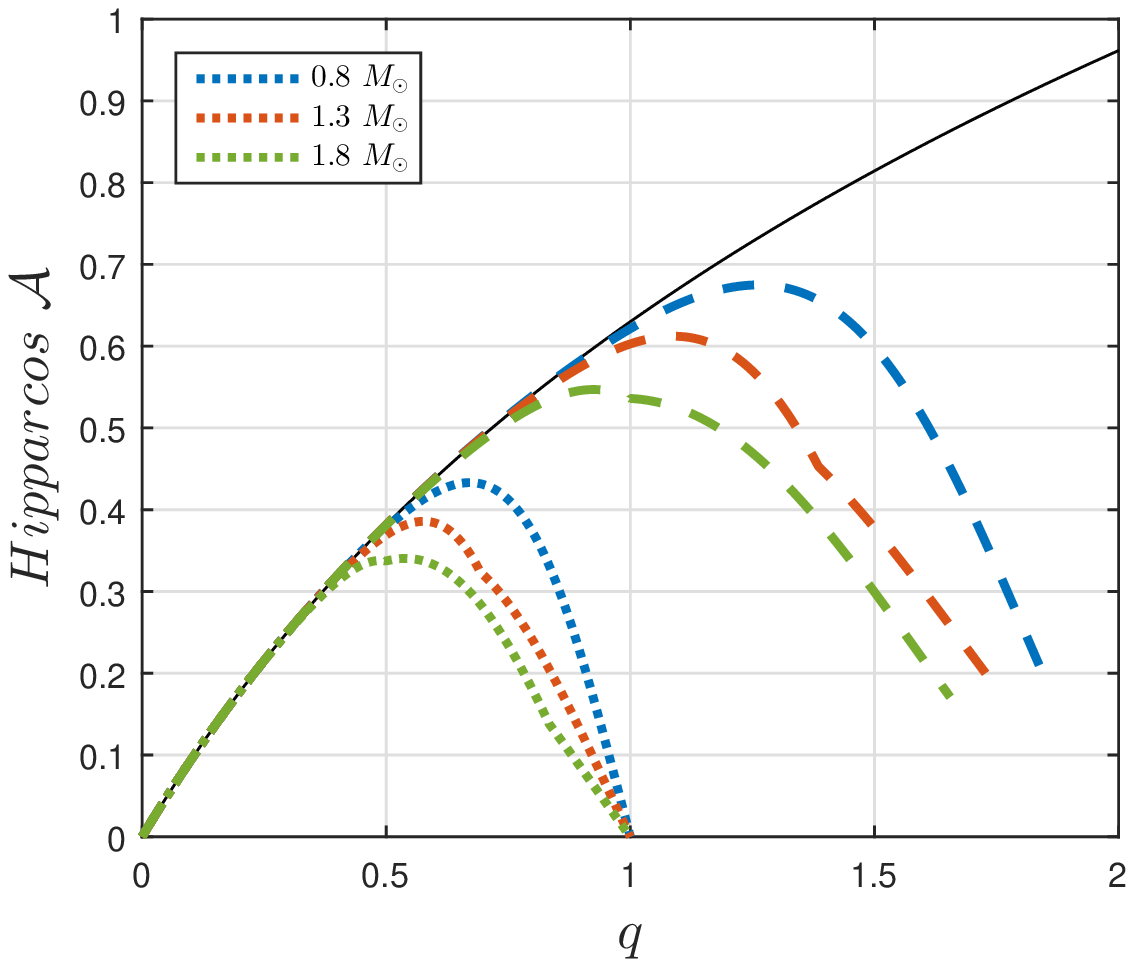}	}
	\caption{ 
AMRF as a function of $q$ for three cases (see text)---$\mathcal A_1$ (a thin continuous line for non-luminous secondary), $\mathcal A_{_{\rm MS}}$ (thick dotted line for an MS secondary) 
and $\mathcal A_{\rm triple}$ (dashed line for hierarchical triple system).
These functions are plotted for $0.8 M_{\odot}$ (blue), $1.3 M_{\odot}$ (red) and $1.8 M_{\odot}$ (green) primaries.
The functions were derived by using equations (A3)--(A5).
}	
\label{fig:HipparcosAms}
\end{figure}
}

\section{Comparison with the SB9 catalogue}
\label{APPENDIX: SB9_validation}
{

\begin{table}
	\begin{tabular}{l c c c}
		\hline 
		\hline
HIP & $\mathcal{A}$ & $\textrm{K}_{\rm ast}$ & $K_1$ \\
    &               & $[km/s]$           & $[km/s]$\\
		\hline
		\hline
5336 & $0.1892 \pm 0.0064$ & $2.37 \pm 0.19$ & $2.13 \pm 0.11$  \\ 
10723 & $0.334 \pm 0.05$ & $15.9 \pm 4.4$ & $19.264 \pm 0.006$  \\ 
37279 & $0.317 \pm 0.011$ & $1.74 \pm 0.1$ & $1.7$  \\ 
45075 & $0.183 \pm 0.013$ & $4.51 \pm 0.46$ & $3.9 \pm 0.04$  \\ 
59750 & $0.342 \pm 0.018$ & $8.13 \pm 0.72$ & $7.71 \pm 0.12$  \\ 
63742 & $0.356 \pm 0.052$ & $11 \pm 3.4$ & $13.03 \pm 0.1$  \\ 
67927 & $0.318 \pm 0.016$ & $8.21 \pm 0.65$ & $8.4$  \\ 
68682 & $0.389 \pm 0.016$ & $6.37 \pm 0.56$ & $6.78 \pm 0.14$  \\ 
70857 & $0.424 \pm 0.029$ & $13.4 \pm 4.2$ & $13.61 \pm 0.05$  \\ 
72848 & $0.357 \pm 0.026$ & $16.5 \pm 1.9$ & $18.913 \pm 0.076$  \\ 
75379 & $0.343 \pm 0.084$ & $13.6 \pm 5.8$ & $14.175 \pm 0.036$  \\ 
75695 & $0.356 \pm 0.017$ & $7.4 \pm 2.9$ & $9.2$  \\ 
82860 & $0.345 \pm 0.036$ & $19.4 \pm 3.9$ & $17.16 \pm 0.004$  \\ 
89808 & $0.298 \pm 0.055$ & $15 \pm 3.8$ & $15.5$  \\ 
		\hline
		\hline 
	\end{tabular} 
\caption{Comparison between our derived semi-amplitudes,
$\textrm{K}_{\rm ast}$, based on the astrometric orbit, and the observed ones, 
$K_1$, for the SB1 systems. The SB9 catalogue does not include errors for $K_1$ for three systems.}
	\label{table:KvsK}
\end{table}

\begin{table}
	\begin{tabular}{l c c c}
		\hline 
		\hline
		HIP & $\mathcal{A}$ & $q_{ast}$ & $q$ \\
		\hline
		\hline
7078 & $0.329 \pm 0.042$ & $0.706 \pm 0.047$ & $0.7524 \pm 0.0019$  \\ 
10644 & $0.334 \pm 0.079$ & $0.78 \pm 0.11$ & $0.888 \pm 0.033$  \\ 
20087 & $0.332 \pm 0.07$ & $0.65 \pm 0.29$ & $0.81 \pm 0.057$  \\ 
65135 & $0.294 \pm 0.044$ & $0.739 \pm 0.077$ & $0.758 \pm 0.042$  \\ 
75389 & $0.385 \pm 0.045$ & $0.787 \pm 0.083$ & $0.787 \pm 0.011$  \\ 
89937 & $0.366 \pm 0.014$ & $0.685 \pm 0.02$ & $0.711 \pm 0.015$  \\ 
101382 & $0.417 \pm 0.055$ & $0.74 \pm 0.31$ & $0.78622 \pm 0.00058$  \\ 
111170 & $0.397 \pm 0.05$ & $0.66 \pm 0.26$ & $0.546 \pm 0.018$  \\ 
		\hline
		\hline 
	\end{tabular} 
	\caption{Comparison between our derived mass ratio based on the astrometric orbit  and the observed ones for the SB1 systems. }
	\label{table:qvsq}
\end{table}

Table~\ref{table:KvsK} compares the expected RV semi-amplitude,  based on the \hipparcos\ orbital elements and our derived $M_1$ and $q$, with $\textrm{K}_1$, the actual observed semi-amplitude RV reported in SB9 for the SB1 systems. 
The errors on $\textrm{K}_\textrm{ast}$ were calculated using the errors on the period, primary mass, inclination angle and derived mass ratio. Only 14 systems had well defined $\textrm{K}_\textrm{ast}$, with small enough errors, that the comparison is meaningful.

Table~\ref{table:qvsq} compares the expected mass ratio  based on the \hipparcos\ orbital elements with mass ratio derived from the SB2 orbits of SB9 catalogue.
}

\end{appendix}
\label{lastpage}
\end{document}